\RequirePackage{lineno}
\documentclass[%
 reprint,
twocolumn,
showpacs,preprintnumbers,showkeys,
 amsmath,amssymb,
 aps,
prc,
floatfix,
]{revtex4}

\usepackage{graphicx}
\usepackage{dcolumn}
\usepackage{bm}
\usepackage{amssymb}
\usepackage{float}
\usepackage[caption = false]{subfig}
\usepackage{hyperref}

\begin{document}


\title{Elliptic flow of $\phi$-meson at intermediate $p_{T}$: Influence of mass versus quark number}
\author{Subikash Choudhury}
\email{subikash.choudhury@cern.ch}
\author{Debojit Sarkar}
\author{Subhasis Chattopadhyay}
\email{sub@vecc.gov.in}
\affiliation{Variable Energy Cyclotron Centre, HBNI, 1/AF Bidhan Nagar, Kolkata 700064, India}
\date{\today}


\begin{abstract}
We have studied elliptic flow ($v_{2}$) of $\phi$-mesons in the framework of  a multi phase transport (AMPT) model
at LHC energy. In the realms of AMPT model we observe $\phi$-mesons at intermediate transverse momentum ($p_{T}$) deviate
from the previously observed (at RHIC) particle type grouping of $v_{2}$ according to the number of quark content i.e, baryons
and mesons. Recent results from the ALICE Collaboration have shown that $\phi$-meson and proton $v_{2}$ has a similar trend,
possibly indicating that particle type grouping might be due to the mass of the particles and not the quark content.
A stronger radial boost at LHC compared to RHIC seems to offer a consistent explanation to such observation.
However, recalling that $\phi$-mesons
decouple from the hadronic medium before additional radial flow is build-up in the hadronic phase,
similar pattern in $\phi$-meson and proton $v_{2}$ may not be due to radial flow alone.
 Our study reveals that models incorporating $\phi$-meson production from $K\bar{K}$ fusion in the hadronic
rescattering phase also predict a comparable magnitude of $\phi$-meson and proton $v_{2}$
particularly in the intermediate region of $p_{T}$. Whereas, $v_{2}$ of $\phi$-mesons created in 
the partonic phase is in agreement with quark-coalescence motivated baryon-meson grouping of hadron $v_{2}$.
This observation seems
to provide a plausible alternative interpretation for the apparent mass-like behaviour of $\phi$-meson $v_{2}$.
We have also observed a violation of hydrodynamical mass ordering between proton and $\phi$-meson $v_{2}$
further supporting that $\phi$-mesons are negligibly affected by the collective radial flow in the hadronic
phase due to the small in-medium hadronic interaction cross sections.
\end{abstract}




\maketitle

\section{Introduction}
The primary objective of heavy ion collisions at ultra relativistic energy is to create and characterize
a novel form of QCD matter consisting of strongly interacting
 and de-confined state of quarks and gluons, the Quark Gluon Plasma (QGP) \cite{intro_1,intro_1a}. 
Dedicated experiments were designed at RHIC and LHC to search for evidences that 
ensure formation of such new state of matter and study its properties. 
One of the key observables, particularly
sensitive to the early stage  dynamics of the collision and hence to the formation of QGP is the elliptic 
flow coefficient $v_{2} = \langle\mathrm{cos [2(\varphi - \Psi_{RP} )]}\rangle$ \cite{intro_2, intro_3, intro_4}.
It quantifies event and particle averaged anisotropy
in the azimuthal ($\phi$) distribution of the particles relative to reaction plane angle ($\Psi_{RP}$) \cite{intro_5}. 

It is generally perceived that in non-central collisions,
the anisotropic emission of final state particles results from the difference
in the pressure gradient in a spatially anisotropic but locally
thermalized system of quarks and gluons. Below $p_{T} < 2$ GeV/c where majority of particles are produced,
this azimuthal anisotropy has been described as a hydrodynamical evolution of strongly interacting QGP  with a nominal 
shear viscosity to the entropy density ratio \cite{intro_6, intro_7, intro_8}
($\eta/s$ extracted is close to AdS/CFT lower bound of 1/4$\pi$). 

Results from RHIC and LHC have revealed that 
$v_{2}$  measured for different particles
as a function of $p_{T}$ exhibits a characteristic mass ordering up to $p_{T}\sim$ 3 GeV/c.
That is, massive particles has less $v_{2}$ and vice-versa at fixed $p_{T}$. Whereas at intermediate $p_{T}$, 3 $\leq p_{T} \leq$ 6 GeV/c,
$v_{2} (p_{T})$ exhibits a flavor ordering i.e,  baryon and meson $v_{2}$ bifurcates \cite{intro_9,intro_10, intro_11}. 
The observed baryon-meson splitting of identified particles $v_{2}$  was found to be compatible
with the models invoking hadronization of a collectively expanding partonic medium
via a mechanism of quark recombination or coalescence \cite{intro_12, intro_13, intro_14, intro_15}. This was further supported by 
the observation of constituent quark number scaling (NCQ) of hadron $v_{2}$,
providing a strong indication towards the onset of the partonic collectivity and the dominance of quark
degrees of freedom at the time of hadronization \cite{intro_16}.

At RHIC energies, baryon-meson difference in $v_{2}$ and NCQ-scaling was taken as a confirmation of quark coalescence
being a plausible mechanism of hadronization at intermediate values of $p_{T}$. 
But at LHC, scaling violation at a level of $\pm$10-20\% and comparable magnitude of 
$v_{2}$ of $\phi$-mesons
and protons in central collisions tend to disfavour coalescence as a relevant particle production mechanism
at this range \cite{intro_10}.
 In hybrid model calculations where partonic and hadronic evolution is modelled by
hydrodynamics and hadronic cascade respectively \cite{intro_17, intro_18}, baryon and meson grouping of $v_{2}$, i.e,  
$v_{2}^{\mathrm{baryons}} > v_{2}^{\mathrm{mesons}}$  at intermediate $p_{T}$
may be understood as a manifestation of
increase in the mean transverse momentum $\langle p_{T} \rangle$ and hence the $p_{T}$-integrated $v_{2}$
values of particles as a function of hadron mass.
Some of these hybrid models also predict upto 30\% increase in the $p_{T}$-averaged $v_{2}$
due to expected rise in the radial boost at LHC when compared
to Au-Au collisions at top-RHIC energy \cite{intro_3, intro_17,intro_18,intro_19, intro_20, intro_21, intro_22}. 
This increase in total transverse boost
could be due to the build-up of additional radial flow in the hadronic phase 
that boosts massive hadrons
to higher $p_{T}$. As the effect is more pronounced for high mass particles,
observed similarity in $\phi$-meson and proton $v_{2}$ appears to be consistent with the increased radial
flow in central A-A collisions at LHC relative to RHIC.
Further studies on the spectral shapes of proton and $\phi$-meson has revealed that in
central collisions $(p + \bar{p}) / \phi$ ratio is independent of $p_{T}$ upto 3-4 GeV/c.
The flat $p_{T}$-dependence of  $p/\phi$ ratio is seen to be in agreement with hydrodynamical
calculations, suggesting the significance of mass over quark number in determining the shape of
$p_{T}$ distributions upto intermediate values of $p_{T}$ \cite{intro_10b}.
Thus, the baryon-meson grouping seems to be congruous with the mass of the particles
rather than the number of quark content \cite{intro_10}.

Generally, those particles which suffer less interactions in the hadronic phase are
often termed as better probes of partonic phase of heavy ion collisions
and may also be sensitive to the particle production mechanism. The hadronic interaction cross section of $\phi$-mesons
with non-strange hadrons because of the OZI-suppression rule is rather small \cite{intro_23, intro_24}. 
Consequently, $\phi$-mesons are not expected to undergo substantial 
rescattering in the late hadronic
phase and decouple from the medium earlier that their non-strange counter parts \cite{intro_25, intro_26}. 
Fact that the $\phi$-mesons are weakly coupled to the medium, radial boost developed during hadronic evolution
has less-significant effect on $\phi$-mesons compared to other hadrons of similar masses.
Thus, the elliptic flow of $\phi$-mesons are expected to be more sensitive to the 
partonic stages of collision and shown to have negligibly affected by hadronic interactions \cite{intro_27, intro_28, intro_28b}.

In contrast, recent measurements by the ALICE collaboration have shown a progressive shift in $\phi$-meson $v_{2}$
from meson to baryon band with increasing centrality and interpreted it as a consequence of
pick-up of some additional radial flow in the post-hadronization phase \cite{intro_10}.
However, considering that $\phi$-mesons decouple prior to the
build-up of radial flow in the hadronic phase,  it seems unlikely
to be an effect of radial flow only. It was shown in \cite{intro_29} that the models incorporating $\phi$-meson production in 
the hadronic rescattering stage via $K\bar{K}$ 
fusion predict a higher value of $\phi$-meson $v_{2}$ relative to other mesons. It would be therefore interesting
to test the effect of hadronic interactions on the elliptic flow of $\phi$-mesons which in-turn may 
be useful in resolving the ambiguity over 
the origin baryon-meson grouping of $v_{2}$ at LHC.

Here, using the string melting (SM) version of a multi phase transport model \cite{ampt_1}
we have calculated $v_{2}$ of some selected species of hadrons including $\phi$-mesons
for Pb-Pb collisions at 2.76 TeV. To demonstrate the effect hadronic rescatterings
on $v_{2}$, model simulation has been performed by varying the time of hadronic cascade.
While discussing our results, emphasis has been given to $v_{2}$ of $\phi$-mesons as they are equally massive
as protons and $\Lambda$s but of different quark content. 
We have also investigated whether the $v_{2}$ of $\phi$ mesons developed at
the partonic phase is modified by additional contributions from the hadronic interactions like,
$K\bar{K}\rightarrow$ $\phi$-meson production.

The presentation of this paper is as follows. 
In section II, we briefly discuss about the AMPT model and  processes of $\phi$-meson
production at the partonic and hadronic stage. 
Results from the model calculation illustrating $v_{2} (p_{T})$
of $\phi$-mesons and other hadrons for different 
hadronic evolution time are shown in Section III and finally we summarize our work in section IV.

\section{The AMPT Model}
\subsection{ Brief description of the model}
AMPT is a hybrid transport model that describes different stages of a heavy ion collision at relativistic energies.
This model has four major steps: the initial conditions, the partonic evolution, the hadronization
and finally the hadronic interactions. As initial conditions, AMPT uses 
spatial and momentum distributions of minijet partons and excited soft strings as implemented in the HIJING event generator \cite{ampt_2}.
Then Zhang's parton cascade (ZPC) \cite{ampt_3}
is used to model the partonic evolution charecterized
by two-body parton-parton elastic scattering with parton interaction cross section
obtained from pQCD calculations  as ${\sigma_{p}\simeq 9\pi\alpha_{s}^{2}/2\mu^{2}}$.
Where $\alpha_{s}$ is the QCD coupling constant for strong interactions and 
$\mu$ is the Debye screening mass of gluons in the QGP medium. At the end of the partonic evolution,
a spatial quark coalescence method is implemented to achieve quark-hadron phase transition in the SM version of AMPT.
In this method,
spatially closed quark-antiquark pairs or triplets are recombined to form mesons and baryons, respectively.
Finally, the hadronic interactions are modelled by A Relativistic Transport calculations (ART) \cite{ampt_4}.

In this study, SM version of AMPT has been used to simulate
Pb-Pb collisions with parton scattering cross sections of 1.5 mb and 3 mb
by keeping the strong coupling constant, $\alpha_{s}$, fixed at 0.33
and tuning the Debye screening mass ($\mu$) to 3.22 fm$^{-1}$
and 2.265 fm$^{-1}$, respectively. 
The parameters for the Lund string fragmentation function, i.e,
\begin{equation}
 f(z) \propto (1-z)^{a}exp(-bm_{T}^{2}/z)
\end{equation}

where z denotes the light cone momentum fraction, are kept same as that of the default HIJING values
corresponding to smaller string tension i,e, a=0.5 and b=0.9 GeV$^{-2}$.

\subsection{Production and interactions of $\phi$ mesons}
In SM version of AMPT $\phi$-mesons are dominantly produced in the partonic stage by coalescence 
of a strange($s$) and an anti-strange($\bar{s}$) quark. During the hadronic evolutions, $\phi$-mesons are also generated
from baryon-baryon interaction channels BB$\rightarrow \phi NN$ and baryon-meson interaction channels 
($\pi,\rho$)B$\leftrightarrow$ $\phi$B, where B= $N, \Delta, N^{*}$ \cite{ampt_1}.
Hadronically, $\phi$-mesons are also produced by kaon-antikaon fusion, $K\bar{K}\rightarrow \phi$,
and the production cross section is obtained from the standard Breit-Wigner form \cite{ampt_6}.

In hadronic rescatterings, $\phi$-mesons also scatter elastically with nucleons and other mesons ($\pi,K,\rho$). 
In this model, elastic scattering
cross section for $\phi$-mesons with nucleons and other mesons are set to 8 mb and 5 mb, respectively \cite{ampt_1}.

\section{Results}
Model calculations based on the SM version of AMPT have shown that at top-RHIC energy
the elliptic flow of $\phi$ mesons are negligibly affected by the hadronic interactions.
While proton $v_{2}$ was found to decrease with the increase in hadronic rescattering time,
$v_{2}$ of $\phi$-mesons remain almost un-altered \cite{intro_17, intro_21aa, intro_22a, intro_27, intro_28b}.
Thus at $p_{T} <$ 1-1.5 GeV/c, $v_{2}^{\mathrm{proton}} <  v_{2}^{\phi}$
although $m_{\phi} > m_{\mathrm{proton}}$, implying a violation in the hydrodynamically expected mass ordering.
The predicted breaking of the hydro-inspired mass-ordering was corroborated
by the recent high-statistics measurements of identified particle $v_{2}$ at RHIC \cite{intro_11}.

But a striking difference was noticed at LHC where $v_{2}$ of $\phi$-mesons at intermediate $p_{T}$
differs from the well-known baryon-meson hierarchy as mentioned in the earlier section. The different
trend of $\phi$-meson $v_{2}$ was argued to be an effect stronger radial flow in Pb-Pb collisions at
$\sqrt{s_{NN}} =$ 2.76 TeV. 
Since the earlier measurements at top-RHIC energy have shown that 
$v_{2}$ of $\phi$-mesons remain almost un-affected because of lower interaction rate in the hadronic
medium, we, therefore re-investigate the effect of hadronic rescatterings
on the elliptic flow of $\phi$-mesons at LHC energy by varying the hadronic evolution (cascade) time from
0.6 to 30 fm/c. Higher time for hadronic cascade corresponds to larger hadronic rescattering.
In the figures, the hadronic cascade time of 30 and 0.6 fm/c are referred to as w/ had. rescatt. and 
w/o had. rescatt., respectively. 

\begin{figure}[htbp]
\centering

\includegraphics[scale=0.45,keepaspectratio]{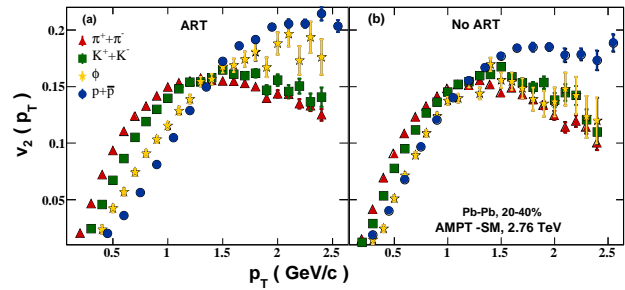}
\caption{[Color online] Elliptic flow parameter $v_{2}$ for pions($\pi^{+} + \pi^{-}$), kaons ($K^{+} + K^{-}$),
phi mesons($\phi$) and protons ($p + \bar{p}$) as a function of transverse momentum calculated from the SM version
of AMPT (a) with hadronic rescattering (b) without hadronic rescattering in 20-40\%
Pb-Pb collisions at $\sqrt{s_{NN}} =$ 2.76 TeV.}
\label{v2pt}
\end{figure}

In Fig.~\ref{v2pt} (a) and (b) we have shown the transverse momentum dependence of elliptic flow  coefficient
( $v_{2} (p_{T})$ ) for pions, kaons, $\phi$-mesons and protons in 20-40\%
Pb-Pb collisions at $\sqrt{s_{NN}} =$ 2.76 TeV from the SM version of AMPT.
The elliptic flow coefficient or $v_{2}$ is obtained by calculating
the 2$^{\mathrm{nd}}$ order Fourier coefficient of azimuthal ($\varphi$) distributions of
final state particles with respect to reaction plane angle ($\Psi_{RP}$), i.e, $v_{2} = < \mathrm{cos 2(\varphi - \Psi_{RP})} >$.
The angular bracket, $<...>$, stands for average over many particles over many events.
For all particles including $\phi$ mesons (decay turned-off), particle identification is done
based on their respective PID or particle identification number in AMPT. At this point it is worth mentioning
that in experiments identification of $\phi$ mesons and its' $v_{2}$ determination differs from the approach 
presented here. First $\phi$-mesons are identified from the invarient
mass distribution of their decay daughters ($\phi \rightarrow K^{+} + K^{-}$) by choosing pairs
within the 3$\sigma$ of $\phi$ mass, followed by $v_{2}$ determination using invarient mass method \cite{intro_28c},etc.
By re-calculating our observable, i.e, $v_{2} (p_{T})$, 
using a different technique (scalar product method), we have checked further whether the choice of a particular method biases
the final conclusion. We found that results obtained from both these methods
are consistent within statistical error. Having established that results are independent of method followed, we now
proceed to discuss their physics implications.

Figure.~\ref{v2pt}(a) represents flow coefficient calculated with hadronic rescatterings and 
Fig.~\ref{v2pt}(b) shows the same without hadronic rescatterings.
These results show that without hadronic rescatterings (Fig.~\ref{v2pt}(b)) the elliptic flow coefficients ($v_{2} (p_{T})$)
exhibit a charecteristic mass ordering, i.e, $v_{2}^{\pi}(p_{T}) > v_{2}^{K}(p_{T})
> v_{2}^{p}(p_{T}) > v_{2}^{\phi}(p_{T})$ for $m_{\pi} < m_{K} < m_{p} < m_{\phi}$
at low $p_{T}$ but the mass splitting is small. On the other hand, as shown in Fig.~\ref{v2pt}(a)
mass splitting increases as hadronic rescatterings are switched-on and a violation of mass ordering between protons and $\phi$-mesons 
$(v_{2}^{p}(p_{T}) < v_{2}^{\phi}(p_{T})$ albeit, $m_{p} < m_{\phi}$)
below $p_{T}$ 1.5 GeV/c is also observed. 
This violation has been interpreted as an effect of different
hadronic interaction cross sections for protons and $\phi$-mesons. As the interaction cross section
of $\phi$-mesons are much smaller than protons, they decouple from the medium earlier and hence  
$\phi$-mesons are negligibly affected by the collective expansion in the hadronic phase.
In contrary, because of significant hadronic interactions $v_{2}$ for protons becomes smaller than that of the $\phi$-mesons which
eventually leads to the breaking of hydrodynamical mass ordering. A more clear picture of this behavior
can be obtained by studying the ratio of $v_{2}^{\phi}(p_{T}) ~\mathrm{to}~ v_{2}^{p}(p_{T})$ as a function transverse momentum.

\begin{figure}[htbp]
\centering

\includegraphics[scale=0.40,keepaspectratio]{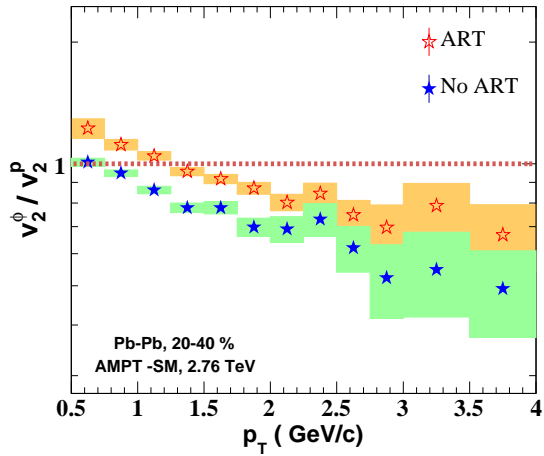}
\caption{[Color online] Ratio of $v_{2}^{\phi}(p_{T}) / v_{2}^{p}(p_{T})$  as a function of transverse momentum calculated from the SM version
of AMPT with hadronic rescattering (open star) and without hadronic rescattering (solid star)  in 20-40\%
Pb-Pb collisions at $\sqrt{s_{NN}} =$ 2.76 TeV. Filled boxes represent statistical uncertainties.}
\label{massviolation}
\end{figure}

It is evident from Fig.\ref{massviolation} that as the hadronic interaction time is increased
from 0.6 fm/c to 30 fm/c (allowing more hadronic rescatterings) the ratio of $v_{2}^{\phi}(p_{T}) /v_{2}^{p}(p_{T})$ 
exceeds unity below 1.5 GeV/c implying breakdown of mass ordering.

Having observed that AMPT-SM with hadronic rescattering has a qualitative agreement with other model calculations
\cite{intro_17, intro_22a} that reasonably describes the identified particles
$v_{2}$ at low $p_{T}$, we now focus on the description of elliptic flow coefficients at the intermediate 
$p_{T}$ region. At RHIC, it was observed that particle production by quark recombination manifest itself in an unique
particle type grouping of $v_{2}$ according to the number of quark content in the intermediate
$p_{T}$ region, i.e,  baryon and meson $v_{2}$ are grouped into two separate branches.

However, at LHC, latest ALICE results show $v_{2}$ of $\phi$-mesons exhibit a different
trend from the particle type grouping. Instead of following the baryon-meson hierarchy, $v_{2}$ values of $\phi$s seem to shifted
towards the baryon band \cite{intro_10}. Our model calculation also reveals that $\phi$-meson $v_{2}$ follow similar
trend as reported by the ALICE Collaboration. As shown in Fig~\ref{v2pt}, $v_{2}$ of $\phi$-mesons
appear to follow the proton (baryon) in presence hadronic rescatterings but falls back on the
meson band when hadronic interactions are turned off. A similar observation was also reported
in this ALICE publication \cite{intro_10}, where it was interpreted as a consequence of
strong radial flow that boosts massive hadrons to higher $p_{T}$.
As $\phi$-mesons and protons have similar masses, they are expected to be boosted equally.

Such observations tend to indicate that baryon-meson grouping could be due to the mass of the particles
rather than the number of constituent quarks. However, recalling that $\phi$-mesons are weakly
coupled to the hadronic medium because of small interaction cross sections and decouples prior to the
build-up of additional radial flow in the hadronic phase, it seems unlikely to be
an effect of radial flow alone.
It was shown in \cite{intro_29} that the models with $\phi$-meson production in 
the hadronic rescattering stage via $K\bar{K}$ 
fusion predict a higher value of $\phi$-meson $v_{2}$ relative to other mesons. It would be therefore interesting
to test the effect of such processes on the elliptic flow of $\phi$-mesons.
 
\begin{figure}[htbp]
\centering

\includegraphics[scale=0.40,keepaspectratio]{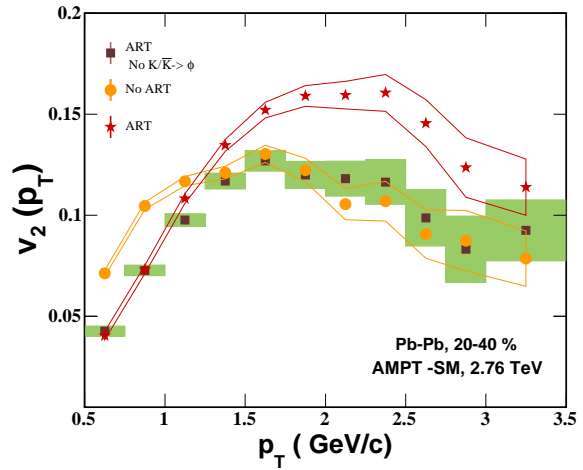}
\caption{[Color online] Transverse momentum dependence $v_{2}$
of $\phi$-mesons calculated from the SM version
of AMPT with hadronic rescattering (solid star), without hadronic rescattering (solid circle)
and with hadronic rescattering but  $K\bar{K}\rightarrow\phi$ forbidden (solid square) in 20-40\%
Pb-Pb collisions at $\sqrt{s_{NN}} =$ 2.76 TeV. Filled boxes and the bands
represent statistical uncertainties}
\label{phimesonv2}
\end{figure}

\begin{figure}[!htp]
\centering

\includegraphics[scale=0.45,keepaspectratio]{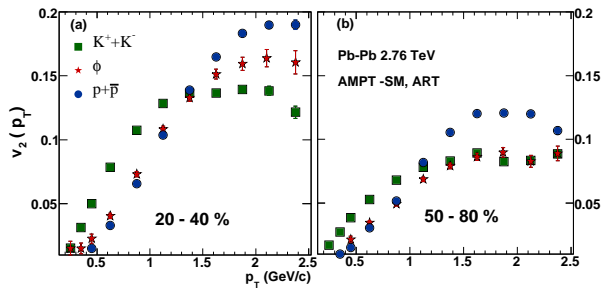}
\caption{[Color online] Elliptic flow parameter $v_{2}$ for kaons ($K^{+} + K^{-}$),
phi mesons($\phi$) and protons ($p + \bar{p}$) as a function of transverse momentum calculated from AMPT SM with
hadronic rescattering in (a) 20-40\% and (b) 50-80\% centrality classes of
Pb-Pb collisions at $\sqrt{s_{NN}} =$ 2.76 TeV.}
\label{v2ptCent}
\end{figure}

In this work we have also analyzed $\phi$-meson $v_{2}$ by turning-off K$\bar{\mathrm{K}}$ coalescence in the hadronic phase.
In Fig.~\ref{phimesonv2} solid star represents $v_{2}$ of inclusive $\phi$-mesons (all $\phi$-mesons produced
in partonic and hadronic phase) and solid square represents $v_{2}$ of $\phi$-mesons excluding
those from the $K\bar{K}$ fusion process (here we call it $\it{primordial}$ $\phi$s).
It is interesting to observe that at the end of hadronic rescattering for 30 fm/c, 
$v_{2}$ of $\it{primordial}$ $\phi$-mesons show no change rather it values at intermediate $p_{T} >$ 1.5 GeV/c coincides
with the results obtained from the model calculation with hadronic re-scatterings turned-off.
Therefore it indicates that $\phi$-mesons regenerated hadronically by $K\bar{K}$ fusion
in the late hadronic stage may be responsible for the observed increase in $v_{2}$ at moderate $p_{T}$. 
But $\it{primordial}$ $\phi$-mesons which are dominantly produced in the partonic phase
are least affected by hadronic interactions and follow quark-recombination expected baryon-meson
grouping. In fact, in peripheral collisions, as shown in the Fig.~\ref{v2ptCent}(b),
even with hadronic rescattering turned-on,
inclusive $\phi$-meson $v_{2}$ is seen to follow meson $v_{2}$ instead of baryon.
\\

\begin{figure}[htbp]
\centering

\includegraphics[scale=0.45,keepaspectratio]{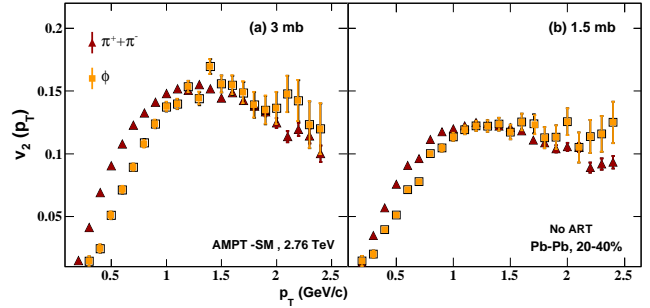}
\caption{[Color online] Transverse momentum dependence of $\phi$-meson and pion $v_{2}$
for 20-40\% Pb-Pb collisions at $\sqrt{s_{NN}} =$ 2.76 TeV. Results obtained from SM version
of AMPT model for parton scattering cross section of (a) 3 mb and (b) 1.5 mb
without hadronic rescatterings.}
\label{phipiv2}
\end{figure}

\begin{figure}[htbp]
\centering

\includegraphics[scale=0.45,keepaspectratio]{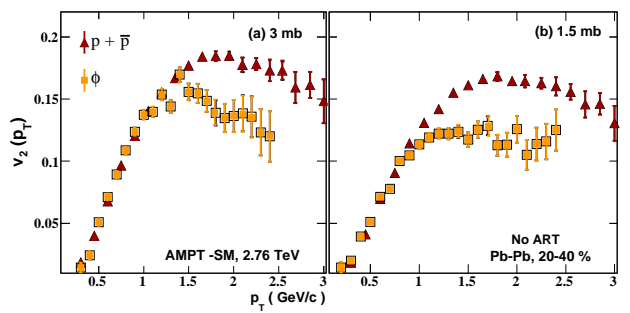}
\caption{[Color online] Transverse momentum dependence of $\phi$-meson and proton $v_{2}$
for 20-40\% Pb-Pb collisions at $\sqrt{s_{NN}} =$ 2.76 TeV. Results obtained from SM version
of AMPT model for parton scattering cross section of (a) 3 mb and (b) 1.5 mb
without hadronic rescatterings.}
\label{phiprv2}
\end{figure}

This could be because of relatively lesser number of
regenerated $\phi$-mesons in peripheral collisions than in central or mid-central collisions
at same $\sqrt{s_{NN}}$. Thus, the apparent mass-like behaviour of $\phi$-meson $v_{2}$ may also be understood as a
consequence of $\phi$-meson regeneration from $K\bar{K}$ fusion.
\\
To further substantiate that $\phi$-meson $v_{2}$ in AMPT is consistent with quark number and not mass, we compare $v_{2}$ of 
$\it{primordial}$ $\phi$-mesons with pions and protons. Results presented in Fig.~\ref{phipiv2} and \ref{phiprv2} 
clearly show, despite mass of $\phi$-meson
being comparable to that of proton (baryon), $\phi$-meson $v_{2}(p_{T})$ at intermediate $p_{T}$ region
exhibit similar flow pattern as that of the lighter mesons irrespective of parton scattering
cross section. Further confirming that in AMPT particle species dependence of the $v_{2} (p_{T})$ is a baryon-meson effect
and not because of the mass of the particle. However, any deviation from the observed pattern may be attributed to the
modification in the spectral shape and/or $v_{2}$ itself
by hadronic interactions in the later stages of collision.

\section{Discussion}
 In summary, we have studied elliptic flow of $\phi$-mesons at low and intermediate ranges of 
transverse momentum for 20-40\% Pb-Pb collisions at 2.76 TeV using a hybrid transport model AMPT.
$\phi$-meson $v_{2}$ has generated lots of interest at LHC since it was observed to deviate from particle type
dependent flow patteren at intermediate $p_{T}$. This observation led to interpretation of baryon-meson ordering of $v_{2}$
as a mass effect rather the quark number. As separate flow patterns for baryons and mesons are naturally
accounted by the hadronization models where hadrons are formed by coalescing quark from a collectively expanding partonic medium,
mass-like flow pattern for $\phi$ mesons would suggest that baryon-meson ordering is simply an interplay between particle mass
and radial flow, which can be explained in the hydrodynamical framework without requiring different hadronization
schemes such as recombination.

However, our model calculation shows that regeneration of $\phi$ during the hadronic phase through hadronic interactions
of K/$\bar{K}$ fusion could be responsible for this apparent mass-like behaviour.
Whereas those created in the partonic phase by $s$-$\bar{s}$ coalescence perfectly follow the baryon-meson grouping.
Inspite of having mass comparable to that of proton, similarity in the $v_{2}(p_{T})$ of $\phi$ and other lighter
mesons ($\pi, K$) further supports that elliptic flow  developed at the partonic phase is inherited by the hadrons
via a mechanism of quark recombination.

At low $p_{T}$, violation in the traditional hydrodynamic mass ordering between proton and $\phi$-meson $v_{2}$ is observed.
This is attributed to small interaction cross section of $\phi$-mesons compared protons resulting in a decrease
in proton $v_{2}$ keeping $\phi$-meson $v_{2}$ almost unaffected during hadronic rescatterings. This observation
is supported by RHIC data for Au-Au collisions at 200 GeV but could not be verified at LHC due to lack of data
below 0.9 GeV/c in $p_{T}$.

\section*{Acknowledgements} 
This research has used resources of the LHC grid computing
facility at Variable Energy Cyclotron Centre, Kolkata.

\end{document}